\documentclass[prd,aps,graphics]{revtex4}
\usepackage{epsfig}

\newcommand{\dd}{{\rm d}}
\newcommand{\bg}{{\vec\gamma}}
\renewcommand{\em}{{\varepsilon_{-1}}}
\newcommand{\eo}{{\varepsilon_{0}}}
\newcommand{\ep}{{\varepsilon_{+1}}}
\newcommand{\Tobs}{{\Theta^{\rm obs}}}
\newcommand{\obs}{{\rm obs}}
\newcommand{\la}{{\ell_1}}
\newcommand{\ma}{{m_1}}
\newcommand{\lb}{{\ell_2}}
\newcommand{\mb}{{m_2}}
\newcommand{\cc}{{\,_2}\widetilde C}
\newcommand{\gc}{{\cal G}}
\newcommand{\fsky}{f_{\rm sky}}
\newcommand{\troisj}[6]{\left(\begin{array}{ccc}
      #1 & #2 & #3 \\
      #4 & #5 & #6\end{array}\right)}
\newcommand{\prodtroisj}[6]{\troisj{#1}{#2}{#3}{#4}{#5}{#6}\troisj{#1}{#2}{#3}{0}{0}{0}}
\begin{document}
\title{Constraints on mode couplings and modulation of the CMB with WMAP data.}

\author{Simon Prunet}
 \email{prunet@iap.fr}
 \affiliation{Institut d'Astrophysique de Paris,
             98 bis, Bd Arago, 75014 Paris, France.}

\author{Jean-Philippe Uzan}
\email{uzan@th.u-psud.fr,uzan@iap.fr}
 \affiliation{Laboratoire de Physique Th{\'e}orique, CNRS--UMR
8627,
         B{\^a}t. 210, Universit{\'e} Paris XI, F-91405 Orsay Cedex,
         France,\\ and \\
         Institut d'Astrophysique de Paris, GReCO,
        CNRS-FRE 2435, 98 bis~Bd~Arago, 75014 Paris, France.}

\author{Francis Bernardeau}
 \email{fbernard@spht.saclay.cea.fr}
 \affiliation{Service de Physique Th{\'e}orique,
         CEA/DSM/SPhT, Unit{\'e} de recherche associ{\'e}e au CNRS, CEA/Saclay
         91191 Gif-sur-Yvette c{\'e}dex}

\author{Tristan Brunier}
\email{brunier@spht.saclay.cea.fr}
 \affiliation{Service de Physique Th{\'e}orique,
         CEA/DSM/SPhT, Unit{\'e} de recherche associ{\'e}e au CNRS, CEA/Saclay
         91191 Gif-sur-Yvette c{\'e}dex}

\vskip 0.15cm

\date{5 juin 2004}

\begin{abstract}
We investigate a possible asymmetry in the statistical properties
of the cosmic microwave background temperature field and to do so
we construct an estimator aiming at detecting a dipolar
modulation. Such a modulation is found to induce correlations
between multipoles with $\Delta\ell=1$. Applying this estimator,
to the V and W bands of the WMAP data, we found a significant
detection in the V band. We argue however that foregrounds and in
particular point sources are the origin of this signal.
\end{abstract}
 \pacs{{\bf PACS numbers:} }
 \vskip2pc

\maketitle
\section{Introduction}\label{sec_1}

The Wilkinson Microwave Anisotropy Probe (WMAP)
data~\cite{wmap1,wmap2} have raised a number of interrogations
concerning the statistical properties of the temperature field.
While these data globally confirm the standard inflationary
paradigm~\cite{peiris} and the concordance cosmological model,
they exhibit some intriguing anomalies, particularly concerning
the large angular scales. In particular, a huge activity has been
devoted to the study of the low value of the quadrupole and
octopole~\cite{quad1,quad2,quad3,quad4} as well as their
alignment~\cite{tegzal,schwarz}, two effects that appear to be
inconsistent with the standard cosmological model.

Besides, many authors have tried to test the statistical
properties of the temperature field using various methods. For
instance, it was investigated whether the coefficients $a_{\ell
m}$ of the development of the temperature field on spherical
harmonics were independent and Gaussian distributed. While, as
expected from the standard inflationary picture, a $\chi^2$
deviation from Gaussianity seems to be well
constrained~\cite{komatsu}, there have been some claims that the
distribution may not be
isotropic~\cite{schwarz,tarun,copi,hansen,eriksen2,eriksen} or
Gaussian~\cite{ng1,ng2,ng3,ng4}. No real physical understanding of
these measurements have been proposed yet and the origin of these
possible features is still unknown. Some authors have argued in
favor of systematic effects~\cite{hansen} while it was
argue~\cite{eriksen2,ss} that foreground contamination may play an
important role in these conclusions.

From a theoretical point of view, there are many reasons to look
for (and/or constrain) a departure from Gaussianity and/or
isotropy of the CMB temperature field. Mode correlation can be
linked to non-Gaussianity, in particular due to finite size
effects~\cite{BU1,BU2,BU3} or to the existence of some non-trivial
topology of the universe~\cite{topology}. While in the latter
case, one expects to have a complex correlation matrix of the
$a_{\ell m}$, the former leads generically to a dipolar modulation
of the CMB field~\cite{BPU}. Such a modulation induces in
particular correlations between adjacent multipoles
($\Delta\ell=1$). Similar correlations but with $\Delta\ell=2$ may
also be induced by a primordial magnetic field~\cite{b1}. In each
case, the physical model and its predictions indicate the type of
correlations to look for and will drive the design of an adapted
estimator.

Investigation of the correlation properties of $a_{\ell m}$ is
thus important to correctly interpret previous observational
results~\cite{schwarz,tarun,copi,hansen,eriksen2,eriksen}. Two
approaches are thus possible. Either one defines some general
estimators and study whether they agree with a Gaussian and
isotropic distribution (top-down approach) or one sticks to a
class of physical models and construct an adapted estimator
(bottom-up approach). In this article, we follow the second route
and focus to the task of constraining a possible dipolar
modulations of the CMB temperature field, that is correlations
between multipoles with $\Delta\ell=1$.

In Section~\ref{sec_1bis}, we start by some general considerations
on the form of the correlation arising from a dipolar modulation.
We then built an estimator, in Section~\ref{sec_2}, adapted to
these types of correlations. In particular, we cannot use full-sky
data and we will need to cut out some part of the sky. The effect
of such a mask on the correlations will have to be taken into
account and included in the construction of the estimator. We
apply this estimator to the V and W bands of the WMAP data in
Section~\ref{sec_3}. The V band exhibits an apparent detection.
The interpretation of this result will require us to compare
various masks, and in particular to investigate the effect of
point sources on the signal to conclude that they are most likely
its cause.

\section{General considerations}\label{sec_1bis}

As explained in the introduction, we focus on a possible dipolar
modulation of the CMB signal. Thus, we assume that the observed
temperature field can be modelled as
\begin{equation}\label{eq2}
 \Tobs(\bg)=\Theta(\bg)\left[1+ \em Y_{1,-1}(\bg) + \eo Y_{1,0}(\bg) + \ep Y_{1,+1}(\bg)\right]
\end{equation}
where $\Theta$ is the genuine statistically isotropic field and
where $(\em,\eo,\ep)$ are three unknown parameters that
characterizes the direction of the modulation. The modulation has
to be real so that $\eo$ is real and
$\ep=-\em^*\equiv\varepsilon$.

As usual, we decompose the temperature fluctuation in spherical
harmonics as
\begin{equation}\label{eq1}
 \Theta(\bg)=\sum_{\ell=2}^\infty\sum_{m=-\ell}^\ell a_{\ell
 m}Y_{\ell,m}(\bg).
\end{equation}
The coefficients $a_{\ell m}$ are thus given by
\begin{equation}\label{eq3}
 a_{\ell m} = \int\dd^2\bg \Theta(\bg) Y_{\ell,m}^*(\bg).
\end{equation}
$\Tobs$ and $a_{\ell m}^\obs$ are defined and related in the same
way. Since $\Theta$ is supposed to be the primordial, Gaussian and
statistically isotropic, temperature field, its correlation matrix
reduces to
\begin{equation}\label{eq8}
 \left<a_{\ell m}\,a_{\ell'm'}^*\right> = C_\ell
 \delta_{\ell\ell'}\delta_{mm'}.
\end{equation}
A modulation of the form (\ref{eq2}) implies that the coefficients
$a_{\ell m}^\obs$ develop correlations between multipoles with
$\Delta\ell=1$. Let us illustrate the origin of this correlation.
From Eqs.~(\ref{eq2}) and~(\ref{eq8}), we deduce that
\begin{equation}\label{eq4}
 a_{\ell m}^{\rm obs} = a_{\ell m} +
 \sum_{\ell'=2}^\infty\sum_{m'=-\ell'}^{\ell'}a_{\ell'm'}
 \sum_{i=-1}^{+1} \varepsilon_i \int\dd^2\bg Y_{\ell,m}^*(\bg)Y_{\ell',m'}(\bg)Y_{1,i}(\bg).
\end{equation}
The integral can be easily computed by using the Gaunt formula
[see Eq.~(\ref{A7})] to get
\begin{eqnarray}\label{eq7}
 a_{\ell m}^{\rm obs} &=&a_{\ell m}+
     \sqrt{\frac{3}{4\pi}}\sum_i\varepsilon_i(-1)^m
     \sum_{LM}\,a_{LM}\,\sqrt{(2\ell+1)(2L+1)}
     \prodtroisj{\ell}{L}{1}{-m}{M}{i}.
\end{eqnarray}
Because of the triangular inequality, the Wigner $3j$-symbols are
non zero only when $L=\ell\pm1$ and $M=m- i$ so that $a_{\ell
m}^{\rm obs}$ is in fact a sum involving $a_{\ell m}$ and
$a_{\ell\pm 1 m- i}$. It follows that it will develop
$\ell-(\ell+1)$ correlations that can be characterized by the two
quantities
\begin{eqnarray}
 D_{\ell m}^{(0)}&\equiv& \left<a_{\ell m}^\obs\,a_{\ell+1m}^{\obs*}\right>,\label{eq9}\\
 D_{\ell m}^{(1)}&\equiv& \left<a_{\ell m}^\obs\,a_{\ell+1m+1}^{\obs*}\right>\label{eq10}
\end{eqnarray}
which will be non zero respectively as soon as $\eo$ or
$\varepsilon$ are non zero. Using the expression (\ref{eq7}) and
the property (\ref{eq8}) of the primordial field, we deduce that
\begin{equation}\label{eq12}
 D_{\ell m}^{(0)} = \eo\,\sqrt{\frac{3}{4\pi}}\,\frac{\sqrt{(\ell+1)^2 -
 m^2}}{\sqrt{(2\ell+1)(2\ell+3)}}\,
 \left(C_\ell + C_{\ell+1}\right)
\end{equation}
\begin{equation}\label{eq13}
 D_{\ell m}^{(1)} = \sqrt{\frac{3}{4\pi}}\,
 \sqrt{\frac{(\ell+2+m)(\ell+1+m)}{(2\ell+1)(2\ell+3)}}\,
 \left[C_\ell +C_{\ell+1}\right]\frac{\varepsilon^*}{\sqrt{2}}.
\end{equation}
Interestingly, these forms indicate how to sum the $D_{\ell m}$ in
order to construct an estimator. This construction will be
detailed in the following section.

\section{Mathematical construction of the estimator}\label{sec_2}

The previous analysis is illustrative but not suitable to be
applied on real data. In particular these data will not be full
sky and we have to take into account the effect of a mask (see
e.g. Ref.~\cite{mask}). Such a mask, that arises in particular
because of the galactic cut, will induce correlations in the
coefficients $a_{\ell m}^\obs$ that are described in
\S~\ref{seca2}. We design the mask in order to protect the
correlations that originate from the modulation (\S~\ref{seca3}
and \S~\ref{seca1}) and finish by presenting the construction of
our estimator in the most general case (\S~\ref{seca4}).

\subsection{Mask effects}\label{seca2}

The temperature field is observed only on a fraction of the sky.
We thus have to mask part of the map so that the temperature field
is in fact given by
\begin{equation}\label{eq20}
 \Tobs(\bg)=\Theta(\bg)\left[1+ \sum_{i=-1}^1 \varepsilon_i Y_{1,i}(\bg)\right]
 W(\bg)
\end{equation}
where $W(\bg)$ is a window function, referred to as mask,
indicating which part of the sky has been cut. We decompose
$W(\bg)$ in spherical harmonics as
\begin{equation}\label{eq21}
 W(\bg)=\sum_{\ell m}\,w_{\ell m}\,Y_{\ell m}(\bg).
\end{equation}
$W(\bg)$ being a real valued function, it implies that $w_{\ell
m}^*=(-1)^mw_{\ell-m}$. We deduce from Eqs.~(\ref{eq20})
and~(\ref{eq3}) that
\begin{equation}\label{eq22}
 a_{\ell m}^{\rm obs} = \widetilde a_{\ell m}
 +\sum_i\varepsilon_iA^{(i)}_{\ell m}
\end{equation}
where $\widetilde a_{\ell m}$ are the coefficients of the masked
primordial temperature field
$\widetilde\Theta(\bg)=\Theta(\bg)W(\bg)$,
\begin{equation}\label{eq22b}
 \widetilde a_{\ell m}= \sum_{\la\ma}\,
 a_{\la\ma}\sum_{\lb\mb}\,w_{\lb\mb}
 \int\dd^2\bg\, Y_{\la\ma}(\bg)Y_{\lb\mb}(\bg)
 Y_{\ell,m}^*(\bg)
\end{equation}
and the effects of the modulation are encoded in the correction
\begin{equation}\label{eq22c}
 A^{(i)}_{\ell m}= \sum_{\la\ma}\,
 a_{\la\ma}\sum_{\lb\mb}\,w_{\lb\mb}
 \int\dd^2\bg\, Y_{\la\ma}(\bg)Y_{\lb\mb}(\bg)Y_{1i}(\bg)
 Y_{\ell,m}^*(\bg).
\end{equation}

Interestingly, $\widetilde a_{\ell m}$ can be shown to be obtained
from $a_{\ell m}$ by the action of a Kernel $K_{\ell m}^{\la\ma}$
\begin{equation}\label{eq23a}
 \widetilde a_{\ell m} = \sum_{\la\ma}\,a_{\la\ma}K_{\ell
 m}^{\la\ma}.
\end{equation}
This kernel is defined by
\begin{eqnarray}
 K_{\ell m}^{\la\ma}&\equiv& \sum_{\lb\mb}\,w_{\lb\mb}
 \int\dd^2\bg\, Y_{\la\ma}(\bg)Y_{\lb\mb}(\bg)
 Y_{\ell,m}^*(\bg)\nonumber
\end{eqnarray}
and can be explicitly computed by using the integral (\ref{A7}) to
obtain
\begin{eqnarray}\label{eq23}
 K_{\ell m}^{\la\ma}
 &=& (-1)^m\sum_{\lb\mb}\,w_{\lb\mb}
   \sqrt{\frac{(2\la+1)(2\lb+1)(2\ell+1)}{4\pi}}
   \prodtroisj{\la}{\lb}{\ell}{\ma}{\mb}{-m}.
\end{eqnarray}

The contribution arising from the modulation can be computed by
using the integral (\ref{A7}) to get
\begin{equation}\label{eqalm}
 A^{(i)}_{\ell m} = \sqrt{\frac{3}{4\pi}}
   (-1)^m\sum_{LM}\,\widetilde a_{LM}\,\sqrt{2\ell+1}\sqrt{2L+1}
   \prodtroisj{L}{1}{\ell}{M}{i}{-m}.
\end{equation}
One can check that the relation (\ref{eq4}) obtained without
taking into account the effects of the mask still holds if one
replaces $a_{\ell m}$ by $\widetilde a_{\ell m}$. The
complications arise from the fact that $\widetilde a_{\ell m}$
does not satisfy the property (\ref{eq8}) because of the action
(\ref{eq23a}) of the Kernel.

\subsection{Choice of the mask and properties of the masked
quantities}\label{seca3}

We now need to specify the form of the mask. First, let us note
that when $W(\bg)=$constant for all $\bg$ then one trivially
recovers that $\widetilde a_{\ell m}=a_{\ell m}$ because
$W=w_{00}Y_{00}$ so that
$$
K_{\ell m}^{\la\ma}
=\frac{w_{00}}{\sqrt{4\pi}}\delta_{\ell\la}\delta_{m\ma}.
$$

\begin{figure}[t]
  \centerline{\epsfig{figure=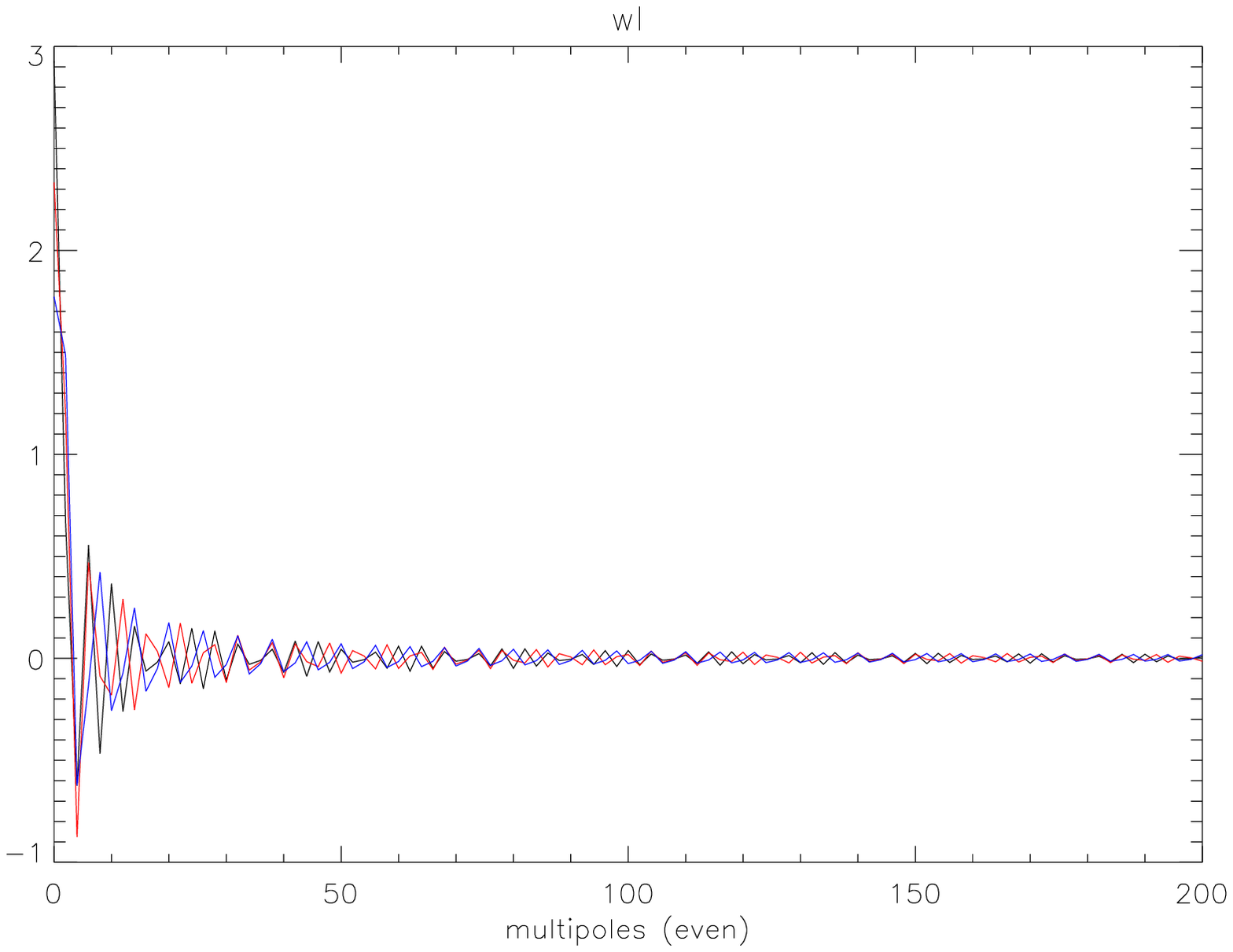,width=8cm}
  \epsfig{figure=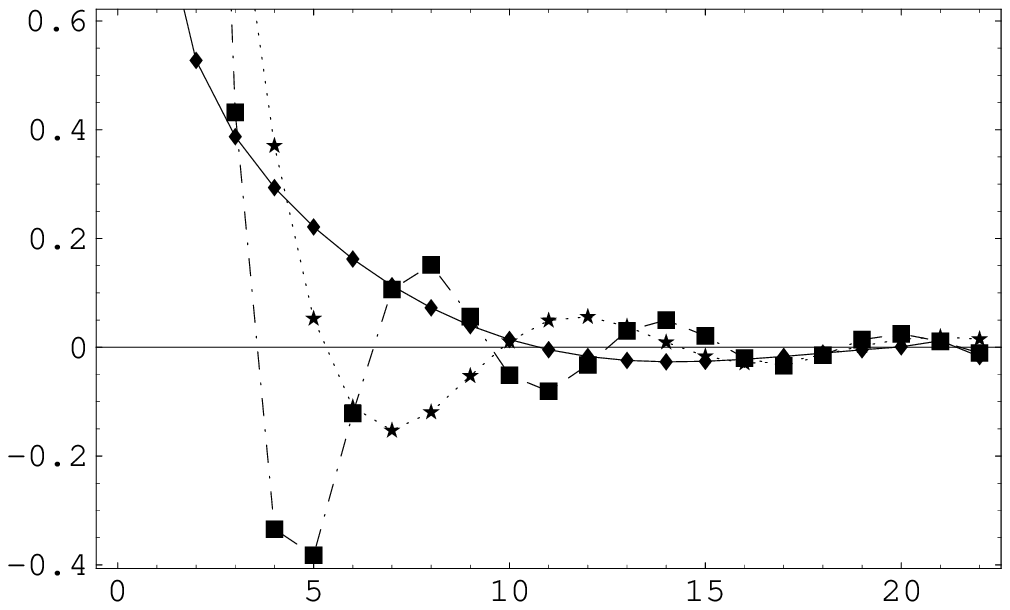,width=8cm}}
  \caption{(left) The coefficients of the decomposition of the mask function
    on spherical harmonics for different galactic cuts, $\mu_c=10,20$ and 30 deg.
    (respectively black, red and blue lines). Only the
    non-vanishing, that is even, multipoles are plotted. (right) The ratio $w_\ell/w_0$
    decreases
    rapidly and becomes typically smaller than 0.1\% for $\ell$ larger than 20 (plain=10 deg., dot=20 deg.,
    dash-dot=30 deg.).}
  \label{fig.wl}
\end{figure}

Since we are looking for $\ell-(\ell+1)$ correlations, we would
like to design a mask that does not involve the same correlations
for $\widetilde a_{\ell m}$ and that is not $m$-dependent. A
solution is to impose that $W(\bg)$ is a function of $\theta$ only
and that it is north-south symmetric, that is
\begin{equation}\label{eqsym}
 W(\bg)=W(\theta),\qquad
 W(\pi-\theta)=W(\theta).
\end{equation}
Since $Y_{\ell 0}(\pi-\theta)=(-1)^\ell Y_{\ell 0}(\theta)$, these
conditions imply that
\begin{equation}\label{w.sym}
 W(\bg)=\sum_{\lb}w_{\lb}\frac{[1+(-1)^\lb]}{2}Y_{\lb 0}(\theta)
 \equiv \sum_{\lb}\hat w_{\lb}Y_{\lb 0}(\theta)
\end{equation}
The simplest example of such a mask is obtained by considering a
function which is constant and vanishes on an equatorial strip of
latitude $\theta_c\in[0,\pi/2]$. This implies that the multipoles
of the mask are given by
\begin{eqnarray}
 w_0 &=& \sqrt{4\pi}\mu_c,\label{mask1}\\
 \hat w_\ell &=& \sqrt{\frac{4\pi}{2\ell+1}}\frac{[1+(-1)^\ell]}{2}
 \left[P_{\ell-1}(\mu_c) - P_{\ell+1}(\mu_c)\right]\label{mask2}
\end{eqnarray}
where $\mu_c=\cos\theta_c$. In particular, it can be seen that
when $\theta_c\rightarrow0$, that is when the size of the mask
vanishes, this mask satisfies
$w_\ell\rightarrow\sqrt{4\pi}\delta_{\ell0}$ when
$\mu_c\rightarrow1$. The function $w_\ell$ is depicted on
figure~\ref{fig.wl} for galactic cuts of 10, 20 and 30 degrees.

The results derived in the following sections are not dependent on
the particular choice of the mask as long as it satisfies the
symmetries (\ref{eqsym}) which ensure that the coefficients of the
mask do not depend on $m$ and vanish for $\ell$ odd (see
Eq.~\ref{w.sym}).

\subsection{Properties of the $\widetilde a_{\ell m}$}\label{seca1}

Whatever the choice of the mask, as long as it satisfies the
properties (\ref{eqsym}), the general expression of the
coefficients $\widetilde a_{\ell m}$ of the decomposition of
$\widetilde\Theta$, are given by
\begin{equation}
 \widetilde a_{\ell m} = \fsky a_{\ell m} +
 (-1)^m\sqrt{2\ell+1}\sum_{\la}\sqrt{\frac{2\la+1}{4\pi}}
 a_{\la m}\sum_{\lb\not=0}
 \sqrt{2\lb+1}\hat w_\lb
 \prodtroisj{\la}{\lb}{\ell}{m}{0}{-m}
\end{equation}
with $\fsky\equiv w_0/\sqrt{4\pi}$ is the fraction of the sky that
is covered. From this expression, we deduce that their 2-point
function is given by
\begin{eqnarray}\label{eqcor}
 \left<\widetilde a_{\ell m}\widetilde a_{\ell'm'}^*\right>
 =\delta_{mm'}\left\lbrace
 C_\ell \fsky^2 \delta_{\ell\ell'} + \fsky
 \left[\gc(\ell,\ell',m)C_\ell + \gc(\ell',\ell,m)C_{\ell'}\right]+
 \sum_\la C_{\la}\gc(\la,\ell,m)\gc(\la,\ell',m)
 \right\rbrace
\end{eqnarray}
where the function $\gc(\ell,\ell',m)$ is defined by
\begin{eqnarray}
\gc(\ell,\ell',m) = (-1)^m\sqrt{(2\ell+1)(2\ell'+1)}
 \sum_{\lb\not=0}\sqrt{\frac{2\lb+1}{4\pi}}\hat w_\lb
 \prodtroisj{\ell}{\lb}{\ell'}{m}{0}{-m}.
\end{eqnarray}
It follows from Eq.~(\ref{eqcor}) that there is no $m$-coupling
arising from the mask (because it has no azimuthal dependence) and
we can define the correlation matrix of the masked temperature
field as
\begin{equation}
 \left<\widetilde a_{\ell m}\widetilde a_{\ell m'}^*\right>
 \equiv \widetilde C_{\ell m}\delta_{mm'}.
\end{equation}
The angular power spectrum of the mask field is then defined as
\begin{equation}
 \widetilde C_{\ell} = \frac{1}{2\ell+1}\sum_{m=-\ell}^\ell \widetilde C_{\ell m}
\end{equation}
and is explicitly given in terms of the primordial angular power
spectrum by
\begin{equation}
 \widetilde C_\ell =
 C_\ell\left[\fsky^2+2\frac{\fsky}{2\ell+1}\sum_{m=-\ell}^\ell \gc(\ell,\ell,m)\right]+ \sum_\la
 \frac{C_\la}{2\ell+1}
 \sum_{m=-\ell}^\ell \gc^2(\la,\ell,m).
\end{equation}

Let us now turn to the $\ell$-$(\ell+1)$ correlators. The first
term in Eq.~(\ref{eqcor}) vanishes. Then, one can check that
$\gc(\ell,\ell+1,m)$ vanishes because the triangular relation of
the Wigner-$3j$ symbols implies that $\ell_2=\pm1$ but for odd
$\lb$ $\hat w_\lb$ vanish. To finish, the contribution of
$\gc(\ell_1,\ell,m)\gc(\ell_1,\ell+1,m)$ in the sum also vanishes
because $\lb$ is even and the sums $\la+\lb+\ell+1$ and
$\la+\lb+\ell$ have to be both even, which is impossible. In
conclusion
\begin{equation}
 \left<\widetilde a_{\ell m}\widetilde a_{\ell+1 m'}^*\right>=
 0.
\end{equation}
As expected from our construction, the mask does not generate
$\ell$-$(\ell+1)$ correlations.

To finish, let us stress that the mask will induce some
$\ell$-$\ell+2$ correlations that can be characterized by
introducing
\begin{equation}
 \cc_{\ell m}\equiv\left<\widetilde a_{\ell m}\widetilde a_{\ell+2
 m}^*\right>,\qquad
 \cc_{\ell}\equiv\frac{1}{2\ell+1}\sum_{m=-\ell}^\ell\cc_{\ell m}.
\end{equation}
Indeed, when $W={\rm Id}$, $\cc_\ell=0$.

\subsection{General construction}\label{seca4}

Starting from the relation~(\ref{eq22}) and the
expression~(\ref{eqalm}), we deduce that the two quantities
defined in Eqs.~(\ref{eq9}-\ref{eq10}) generalize to
\begin{eqnarray}
 D_{\ell m}^{(i)} &=& \sqrt{\frac{3}{4\pi}}\varepsilon_i^*
 \left\lbrace
   (-1)^{\ell + m + 1 + i} \sqrt{\ell + 1}
   \troisj{\ell}{1}{\ell+1}{m}{i}{-m-i}
   \left[ \widetilde C_{\ell m} + \widetilde C_{\ell+1 m+i}
   \right]\right. \nonumber\\
   &&\qquad\qquad
   +(-1)^{\ell+m+i}\sqrt{\ell}\troisj{\ell-1}{1}{\ell}{m+i}{-i}{-m} \cc_{\ell-1 m+i}
   \nonumber\\
   &&\qquad\qquad\left.
   +(-1)^{\ell+m+i}\sqrt{\ell+2}\troisj{\ell+1}{1}{\ell+2}{-m-i}{i}{m} \cc_{\ell
   m}
   \right\rbrace,
\end{eqnarray}
with $i=0,1$, when the mask effects are taken into account. This
expression is defined for $m=-\ell\ldots\ell$ even if $\cc_{\ell-1
m+1}$ is not defined for $m=\ell$ and $m=\ell-1$ and $\cc_{\ell-1
m}$ for $m=\ell$ because the Wigner-$3j$ symbols that multiply
these terms strictly vanish. From this expression, we define
\begin{equation}
 D_\ell^{(i)} \equiv \frac{\sum_{m=-\ell}^\ell (-1)^{\ell+m+1+i}
 \troisj{\ell}{1}{\ell+1}{m}{i}{-m-i}D^{(i)}_{\ell m}}{\sum_{m=-\ell}^\ell (-1)^{\ell+m+1+i}
 \troisj{\ell}{1}{\ell+1}{m}{i}{-m-i}}.
\end{equation}
Now, it can be checked, after some algebra, that
\begin{equation}\label{eq34}
 D_\ell^{(i)} = \sqrt{\frac{3}{4\pi}}\varepsilon_i^*\left[
 \hat C_\ell^{(i)} + 2\, {_2}{\hat C}_\ell^{(i)A}+ 2\,{_2}{\hat
    C}_\ell^{(i)B}
 \right] + {\cal O}(\varepsilon^2)
\end{equation}
where the quantities $\hat C_\ell^{(i)}$, ${_2}{\hat
C}_\ell^{(i)A}$ and ${_2}{\hat C}_\ell^{(i)B}$ have been defined
by
\begin{eqnarray}
 \hat C_\ell^{(i)} &=& \frac{1}{N^{(i)}_\ell}
      \sum_{m=-\ell}^\ell\sqrt{\ell+1)}
      \troisj{\ell+1}{1}{\ell}{m}{i}{-m-i}^2
      \left[\widetilde C_{\ell m}+ \widetilde C_{\ell+1 m+i}
      \right]\\
 {_2}{\hat C}_\ell^{(i)A} &=& -\frac{1}{N^{(i)}_\ell}
      \sum_{m=-\ell}^\ell\sqrt{\ell+2}
      \troisj{\ell}{1}{\ell+1}{m}{i}{-m-i}
      \troisj{\ell+2}{1}{\ell+1}{m}{i}{-m-i}
       {_2}\widetilde C_{\ell m}\\
{_2}{\hat C}_\ell^{(i)B} &=& -\frac{1}{N^{(i)}_\ell}
      \sum_{m=-\ell}^\ell\sqrt{\ell}
      \troisj{\ell}{1}{\ell+1}{m}{i}{-m-i}
      \troisj{\ell-1}{1}{\ell}{m+i}{-i}{-m}
       {_2}\widetilde C_{\ell m}
\end{eqnarray}
with the coefficients $N^{(i)}_\ell$ given by
\begin{equation}
 N^{(i)}_\ell = \sum_{m=-\ell}^\ell
      (-1)^{\ell+m+i}
      \troisj{\ell}{1}{\ell+1}{m}{i}{-m-i}.
\end{equation}

It follows from these results that we can consider the estimator
\begin{eqnarray}\label{eq39}
  E_\ell^{(i)} &=& \frac{1}{N^{(i)}_\ell} \sum_{m=-\ell}^\ell \left(\begin{array}{ccc}
      \ell & 1 & \ell+1 \\
      m & i & -m-i
    \end{array}\right) (-1)^{\ell+m+i}\,
    a_{\ell m}^\obs a_{\ell+1 m+i}^{\obs*}
\end{eqnarray}
that satisfies by construction
\begin{eqnarray}
 \left<E^{(i)}_\ell \right>=D_\ell^{(i)}.
\end{eqnarray}
We will apply this estimator to the WMAP data in the following
sections.

\section{Data analysis}\label{sec_3}

The proposed estimators have been implemented numerically, using
the Healpix \footnote{{\rm http://www.eso.org/science/healpix}}
package for the pixelization and the fast spherical harmonics
transforms, and applied to the co-added data of the WMAP V and W
bands (resp. ${\rm 70\, GHz}$ and ${\rm 90\,GHz}$) where most of
the signal is of cosmological origin. We implemented the
estimators as described by equations~(\ref{eq34}) to (\ref{eq39}).

The quantities $\hat{C}_\ell^{(i)}$, ${}_2\hat{C}_\ell^{(i)A}$ and
${}_2\hat{C}_\ell^{(i)B}$ have been computed using the best fit
LCDM theoretical power spectrum of the WMAP data \cite{verde}, and
were not computed on the data itself to avoid ratios of random
variables. To assess the statistical significance of the measured
values of $\varepsilon_i$, we made $1000$ simulations of WMAP data in
each of the V and W bands according to a sky model with no
modulation.

\begin{figure}[t]
 \centerline{\epsfig{figure=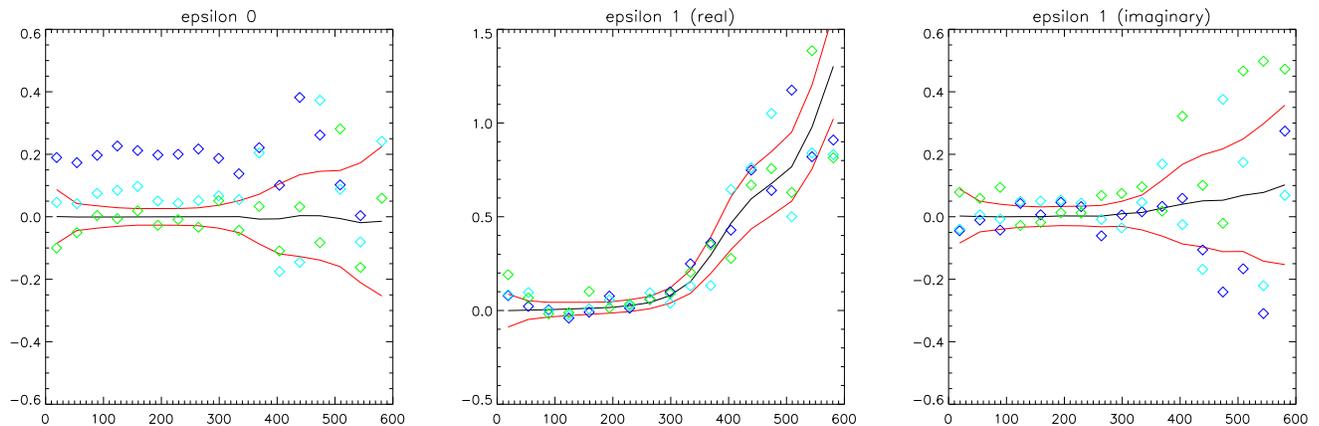,width=18cm}}
 \caption{$D_\ell$ measured on the WMAP data (W band) [green triangles]. Blue
 triangles are the measurements on a simulated map with $\eo=0.2$ and the
 red lines are $1\sigma$ error bars.}
 \label{fres1}
\end{figure}

\begin{figure}[t]
 \centerline{\epsfig{figure=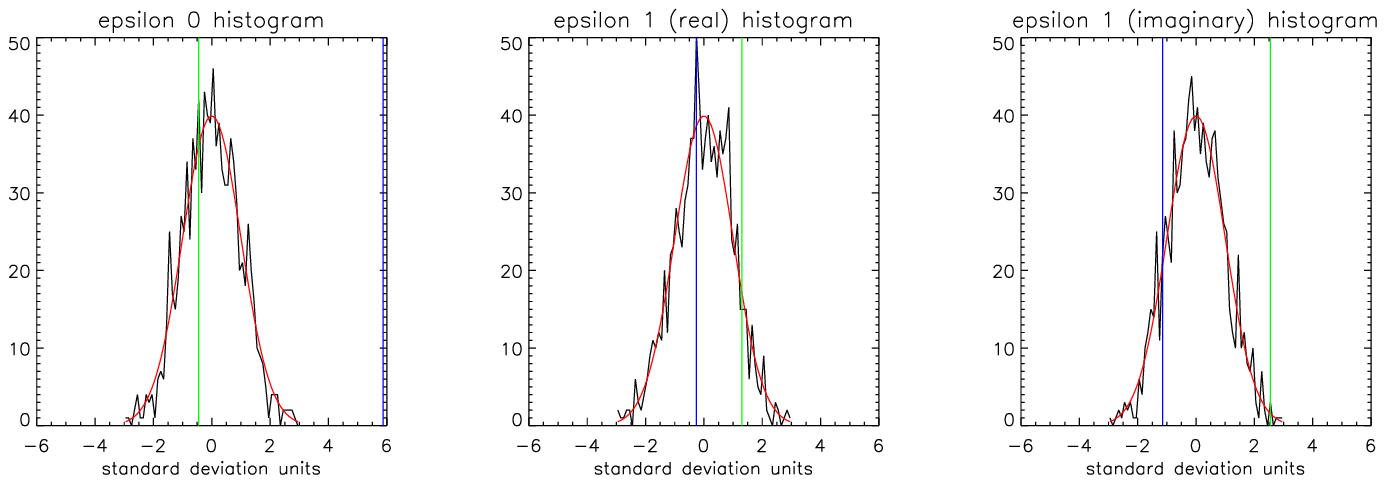,width=18cm}}
 \centerline{\epsfig{figure=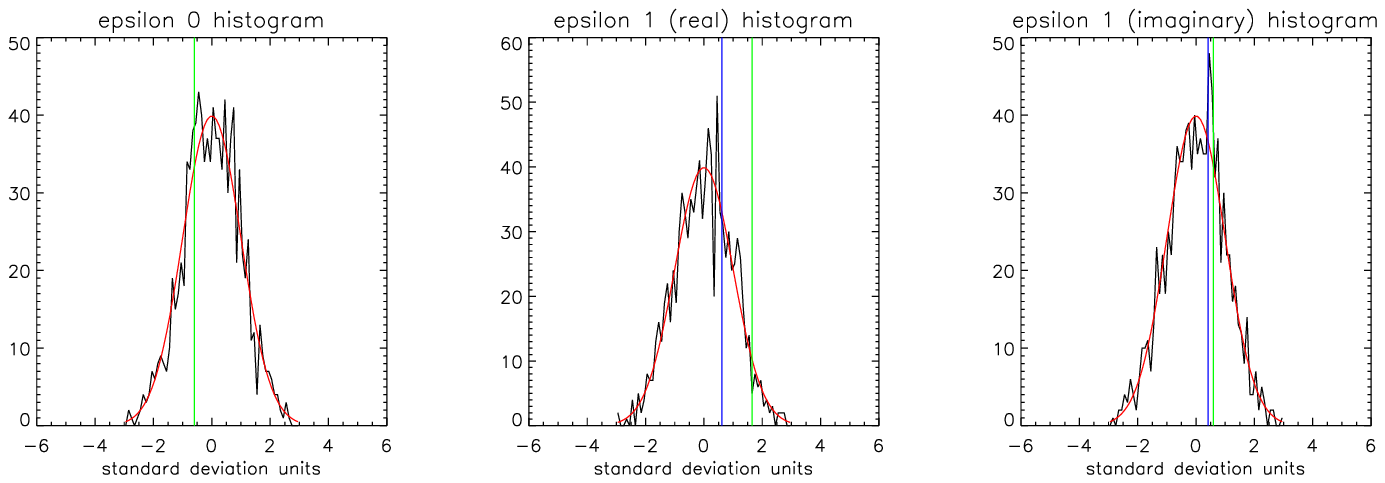,width=18cm}}
 \caption{Comparison of 1000 simulations with the WMAP data. We use the
 W band and sum the multipole between $\ell=20$ and $\ell=100$ (top panel)
 and between $\ell=100$ and $\ell=300$ (bottom panel). The color code is identical
 to the one of Fig.~\ref{fres1} that is the green line correspond to the measurement
 on the WMAP data and the blue line the measurement on a simulation.}
 \label{fres2}
\end{figure}

\begin{figure}[b]
 \centerline{\epsfig{figure=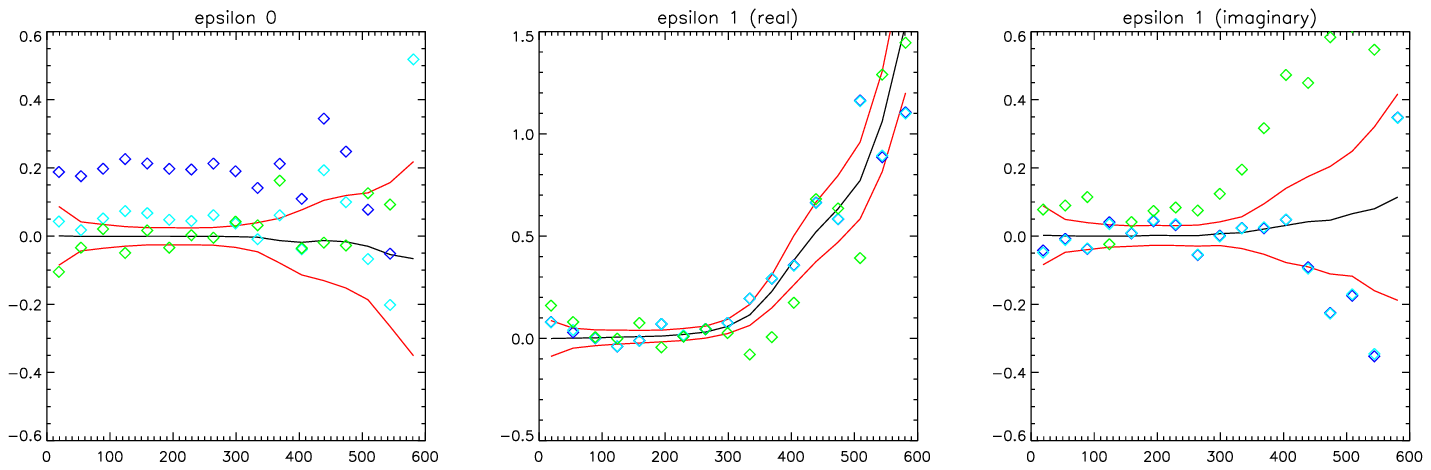,width=18cm}}
 \caption{$D_\ell$ measured on the WMAP data (V band) [green traingles]. Blue
 triangles are the measurements on a simulated map with $\eo=0.2$ and the
 red lines are $1\sigma$ error bars.}
 \label{fig:epsilon_V_35}
\end{figure}

\begin{figure}[b]
 \centerline{\epsfig{figure=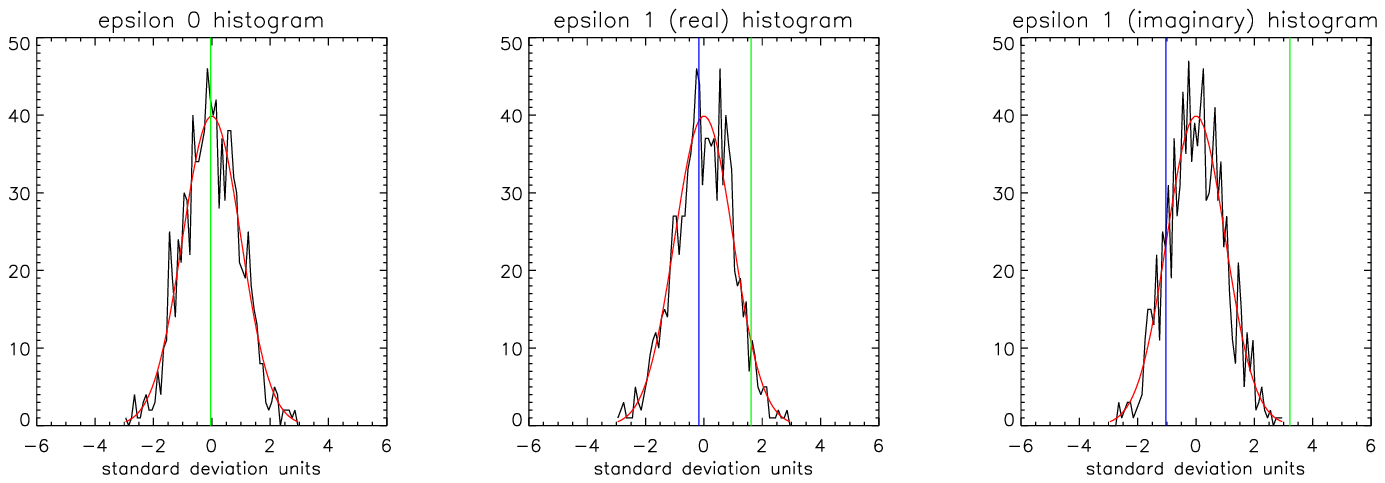,width=18cm}}
 \centerline{\epsfig{figure=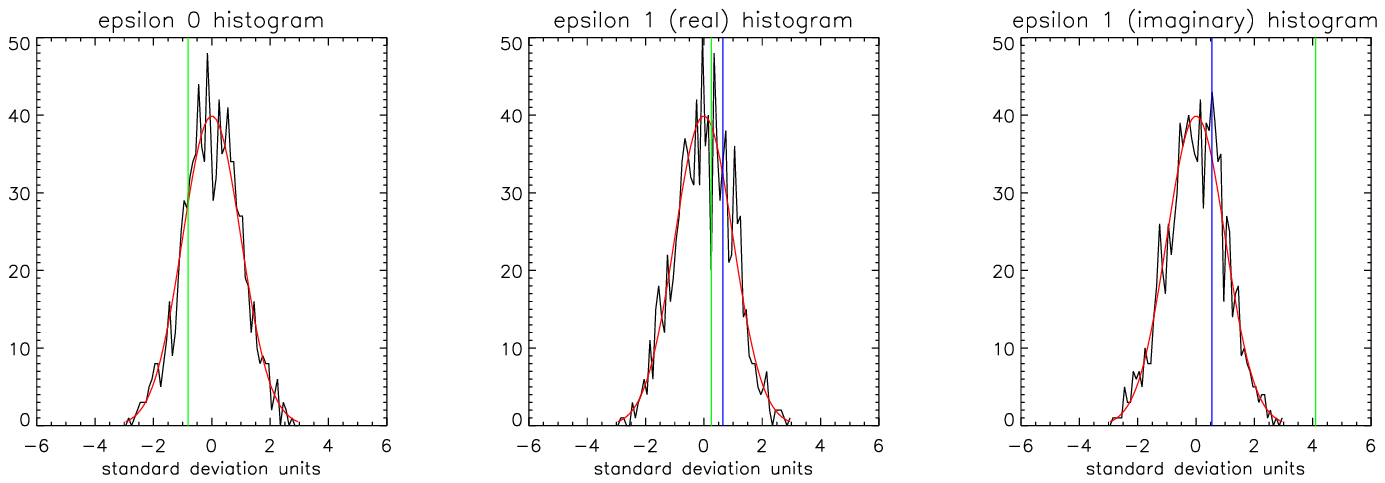,width=18cm}}
 \caption{Comparison of 1000 simulations with the WMAP data. We use the
 V band and sum the multipole between $\ell=20$ and $\ell=100$ (top panel)
 and between $\ell=100$ and $\ell=300$ (bottom panel).}
 \label{fig:histo_V_20_100}
\end{figure}

The results of the analysis of the V and W bands are summarized on
Figs.~\ref{fres1} to~\ref{fig:histo_V_20_100}. Fig.~\ref{fres1}
depicts the measurement of $D_\ell$ on the W band. We sum this
measurement of two bands of $\ell$ (respectively 20-100 and 100-300)
and compare with 1000 simulated WMAP data. We perform the same tests
on the V band (Fig.~\ref{fig:epsilon_V_35}
and~\ref{fig:histo_V_20_100}). The apparent detection in the V band
without clear counterpart in the W band suggest a non cosmological
contamination. Determining its origin requires to performed more
tests.

This contamination can be a priori from two possible sources, Galactic
or extragalactic. To check if the correlations detected in the V band
are of Galactic origin, we apply the same estimator to the half sum
and half difference of the V and W bands, that is
\begin{equation}\label{SD}
 S=\frac{W+V}{2},\qquad
 D=\frac{W-V}{2}
\end{equation}
and repeat the whole procedure on 1000 simulation in each case,
where the simulations contain only CMB and noise according to the
WMAP specifications. The advantage of the half difference of the
bands is that it should (up to calibration errors) eliminate the
CMB signal completely at large scales, hence eliminate the main
source of variance at these same scales, where the Galactic
signals are expected to dominate. Indeed, the power spectra of
Galactic emissions usually scale as $C_\ell \propto
\ell^{-\alpha}$, with $2\leq \alpha \leq 3$ (see e.g.
Ref.~\cite{bg}). The half sum results, summarized in table~I,
are in between those of the V and W bands, which
is coherent with the assumption of the detection being caused by a
foreground source of electromagnetic spectrum different from the
CMB fluctuations.

More importantly, the half difference results do not show a strong
correlation detection at large angular scales, in contradiction
with the assumption of the Galactic foreground contamination being
the source of the detected correlations in the V band.

However, this half difference test does not work that well if the
source contaminants are of extragalactic origin, since the power
spectra of extragalactic foregrounds resemble that of the noise.
In this case, the contamination is expected to increase with
increasing multipole number, which seems to be the case for the V
band (see Figs.~\ref{fig:epsilon_V_35} and
\ref{fig:histo_V_20_100}).

\begin{figure}[t]
  \centerline{\epsfig{figure=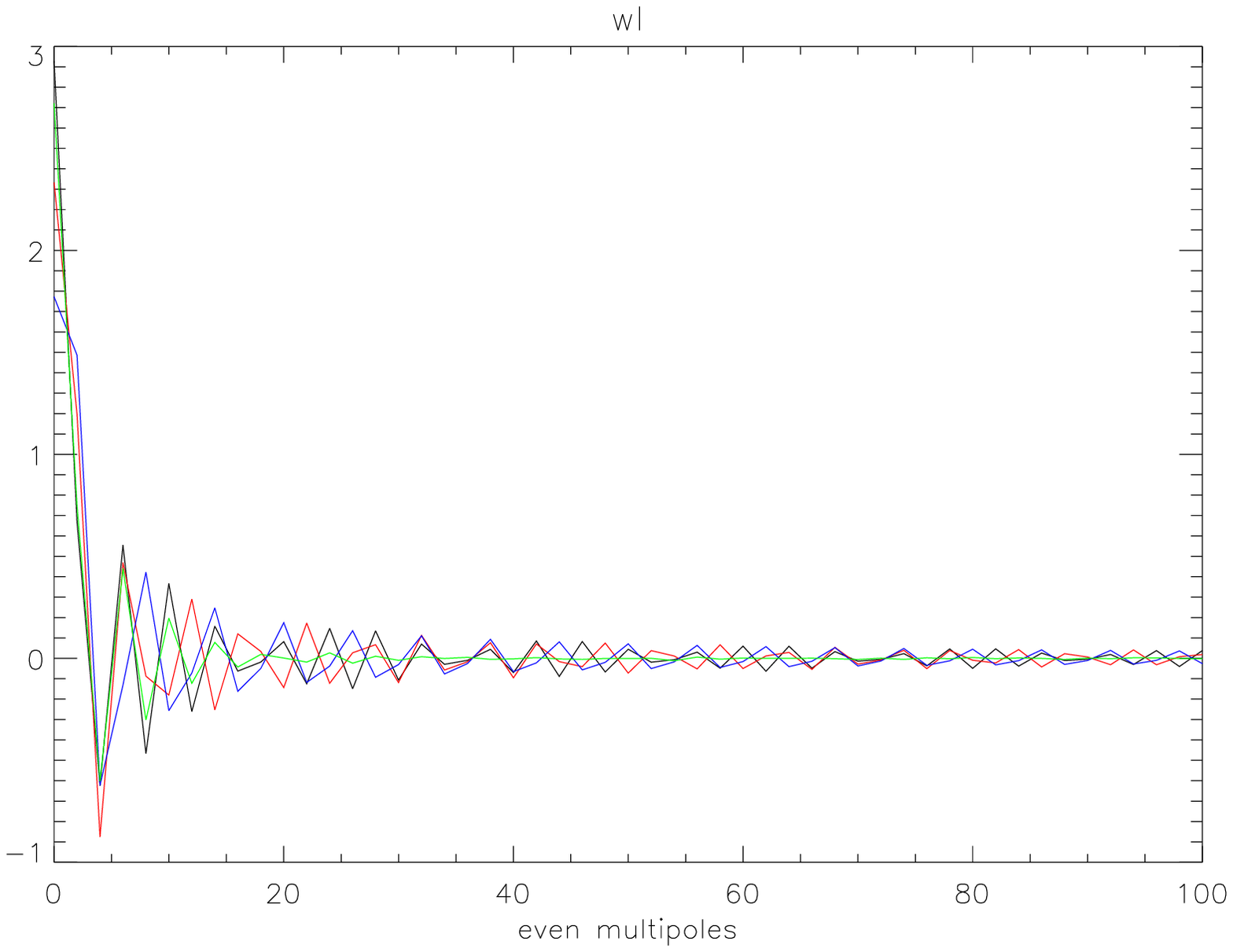,width=16cm}}
  \caption{The coefficients of of the decomposition of the mask function
    on spherical harmonics for different galactic $\mu_c=10,20$ and 30 deg.
    (respectively black, red and blue lines) compared with the ones of the kp0
    mask (green).}
  \label{fig_wchange}
\end{figure}

The difficulty of extragalactic point sources contamination is
that these sources (quasars and active radio-galaxies) are
distributed more or less uniformly across the sky, which renders
their masking by an azimuthally symmetric sky cut impossible.
However, the WMAP team provides with their data sets ``taylor
cuts'' that blank out the resolved point sources of largest flux.
Of course, the dipolar modulation estimators designed in the
preceding sections do not apply stricto sensu to these arbitrary
masks, but one can hope, given the small fraction of sky removed
at high latitude, that the broken symmetry of the mask will be a
small perturbation in the computation of the $\varepsilon$'s, so that
the estimators keep their general validity, up to a possible small
bias (see Fig.~\ref{fig_wchange} for a comparison of the
coefficients $w_\ell$ of the two masks).

\begin{figure}[t]
\centerline{\epsfig{figure=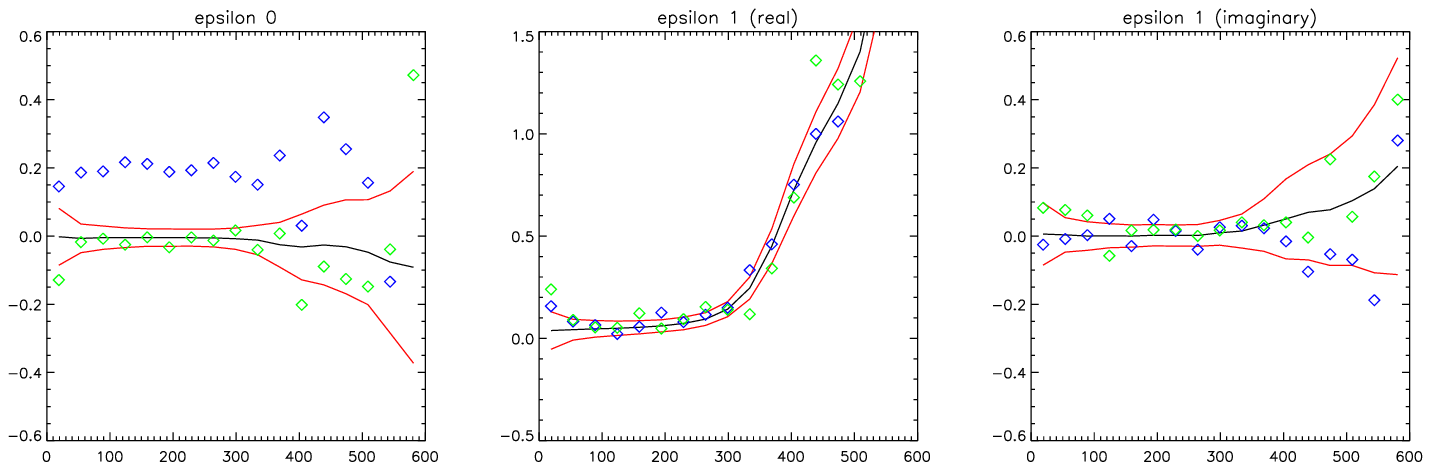,width=18cm}}
 \caption{$D_\ell$ measured on the WMAP data (V band) [green
  triangles], using the taylor
  mask kp0 to blank the main point sources. Blue
  triangles are the measurements on a simulated map with $\eo=0.2$ and the
  red lines are $1\sigma$ error bars.}
  \label{fig:epsilon_Vkp0_35}
\end{figure}

\begin{figure}[t]
 \centerline{\epsfig{figure=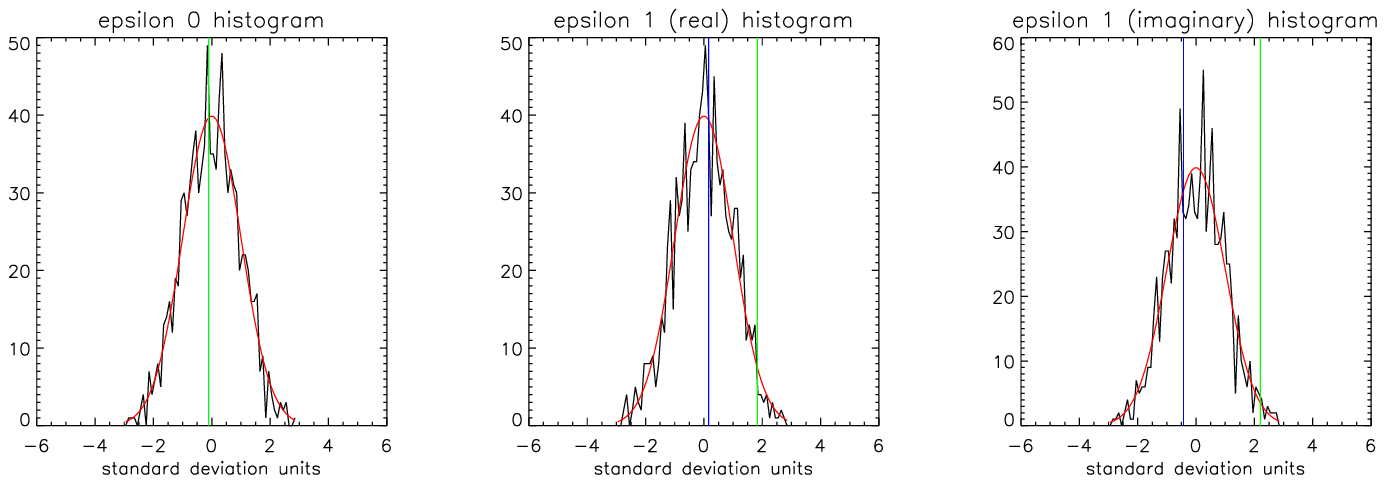,width=18cm}}
 \centerline{\epsfig{figure=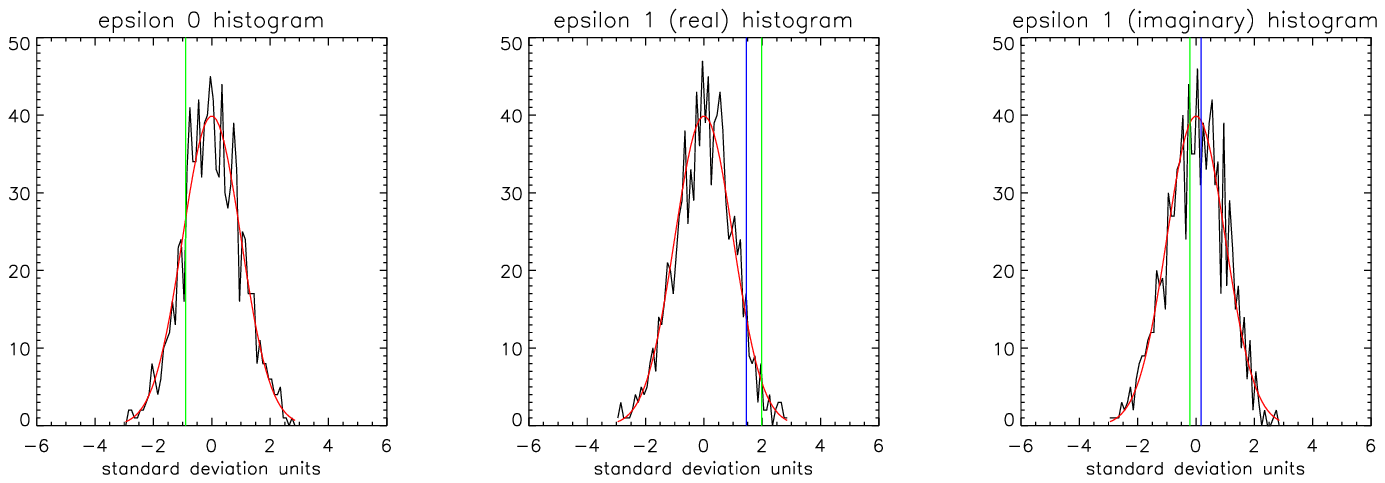,width=18cm}}
 \caption{Comparison of 1000 simulations with the WMAP data. We use the
 V band and sum the multipole between $\ell=20$ and $\ell=100$ (top panel)
 and between $\ell=100$ and $\ell=300$ (bottom panel) using the taylor
mask kp0 to blank the main point sources.}
 \label{kp0_histo}
\end{figure}

This assumption can be checked on a simulated sky with a known
dipolar modulation, where a WMAP V-band noise is added to the
signal, together with the taylor mask. We chose the most
conservative mask provided by the WMAP team (kp0), and applied it
to a simulated sky of known modulation ($\varepsilon_0=0.2$) as
described above, together with the V-band data. The results are
shown in figures~\ref{fig:epsilon_Vkp0_35} and ~\ref{kp0_histo}. Again, the estimators
have been applied to $1000$ simulations of the V band with no
dipolar modulation, with the same kp0 mask applied, to estimate
the statistics of the V-band data results.

Several observations can be made at this point. First, comparing
these results with those obtained in the V-band, but with the
azimuthally symmetric $20^\circ$ cut
(figure~\ref{fig:epsilon_V_35}), one can check that the estimators
give very compatible results for the simulated dipolar modulation
(blue squares). This comforts our assumption that changing the cut
sky to the kp0 mask is a small perturbation for the modulation
estimators.

Secondly, comparing the same figures but this time looking at the
data (green squares), one can see that in the case of the
$20^\circ$ cut there is a large trend at high $\ell$'s in
$\varepsilon_1$ that disappears when the kp0 cut is used. This is
confirmed by the results of table~\ref{table:results} where one
can check that the tentative detections of in the V band using the
simple cut become statistically insignificant when using the kp0
cut.

\section{Discussion and conclusions}\label{sec_4}

In this article we have proposed an estimator designed to detect a
possible modulation of the CMB temperature field, or equivalently
$\Delta\ell=1$ correlations. The effects of cutting part of the
sky were discussed in details and we applied this estimator to the
V and W bands of the WMAP data.

The results of our analysis are summarized in
table~\ref{table:results} which gives the amplitude of the
modulation coefficients on the WMAP data and a corresponding test
case with $\eo=0.2$, $\hbox{Re}(\varepsilon_1) =
\hbox{Im}(\varepsilon_1) = 0$. All values are given in standard
deviation units, estimated on 1000 (signal+noise) simulations in
each case, with no modulations.

While the V band seems to exhibit a marginal detection, further
tests such as the study of the half sum and difference of the
two bands and the effect of point sources have led us to conclude
that this detection should be inferred to the effect of point
sources contamination. In this analysis we have used the kp0 mask
which does not satisfied the symmetries of the mask required for our
estimator to be unbiased. Nevertheless, our estimator seems to be
well suited for the analysis, even with the kp0 mask.

To back up this interpretation we have performed two last tests.
First we added to a simulated CMB map without modulation and with
noise the 208 sources resolved by the WMAP experiment and then
smoothed with the correct beam. Second we added to the same
simulation the 700 circular region that are cut in the analysis of
the V band in the WMAP analysis. Both simulations, while analyzed
as the previous data with an azimuthal mask of 20 deg., exhibit an
excess of signal for $\eo$ and $\hbox{Im}(\varepsilon_1)$ in the
same range of multipoles than obtained on the analysis of the V
and W bands (Figs.~\ref{fres1} and~\ref{fig:epsilon_V_35}).
Interestingly, the signal of $\hbox{Re}(\varepsilon_1)$ is not
affected and is identical to the one of Figs.~\ref{fres1}
and~\ref{fig:epsilon_V_35}. Indeed, the signals have not exactly
the same amplitude as the ones obtained from the analysis of the V
band but they exhibit the same trend on the the same scales. Also,
it has to be stressed that with a cut of 20 deg. the Large
Magellannic Cloud (galactic latitude of 20 deg. and more and
longitude of 0 deg.) and a part of the H2 Ophucius region should
contribute and that we have not included them in the simulations.
This could have enhance the signal.

In conclusion, the set of analysis performed in our study tend to
show that the $\Delta\ell=1$ correlations that appeared in the
analysis of the V and W bands of the WMAP data are due to
foreground contaminations and most likely by point sources. The
direction of the detected modulation will, in that interpretation,
characterize the anisotropy of the distribution of these sources.

\begin{table}
\begin{tabular}{|l||c|c||c|c||c|c|}
\hline
 &\multicolumn{2}{c}{$\eo$}&\multicolumn{2}{c}{Re($\varepsilon$)}&\multicolumn{2}{c}{Im($\varepsilon$)}\\
 & data & test & data & test & data & test \\
 \hline
 W (20-100)  & $\quad$-0.45$\quad$ & $\quad$5.87$\quad$ & $\quad$1.30$\quad$ &
    $\quad$-0.26$\quad$ & $\quad$2.54$\quad$ & $\quad$-1.14$\quad$ \\
 W (100-300) & -0.60 & 16.9 & 1.65 & 0.61  & 0.59 & 0.41  \\
 \hline
 V (20-100)  & -0.04 & 6.00 & 1.61 & -0.17 & {\bf 3.21} & -1.03 \\
 V (100-300) & -0.81 & 17.9 & 0.25 & 0.65  & {\bf 4.10} & 0.54  \\
 V-kP0 20-100)  & -0.11 & 6.12 & 1.83 & 0.16 & 2.20 & -0.42\\
 V-kp0 (100-300)$\,$& -0.89 & 17.4 & 1.98 & 1.45 & -0.22& 0.18\\
 \hline
 S (20-100)  & -0.24 & 6.71 & 1.52 & 0.40 & 2.85 & -0.31 \\
 S (100-300) & -0.64 & 19.3 & 1.15 & 0.57 & 2.16 & 1.35  \\
 \hline
 D (20-100)  & -0.58 & -0.74 & -2.10 & -1.49 & 3.73 & -0.70 \\
 D (100-300) & -0.98 & 0.93  & -0.44 & 0.69  & 2.67 & -0.54 \\
 \hline
\end{tabular}
\caption{Summary of the data analysis performed in this article.
It concerns the two bands V and W, their half sum (S) and half
difference (D). The band V has been analyzed with two masks to
emphasize the effect of the point sources on the result.}
\label{table:results}
\end{table}

\begin{figure}[t]
  \centerline{\epsfig{figure=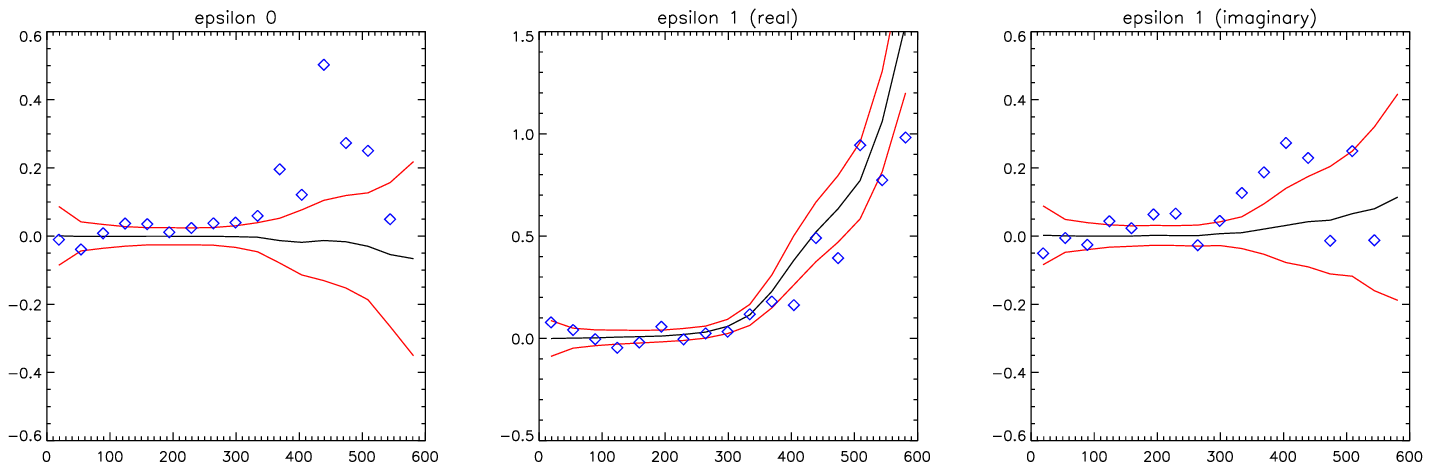,width=18cm}}
  \centerline{\epsfig{figure=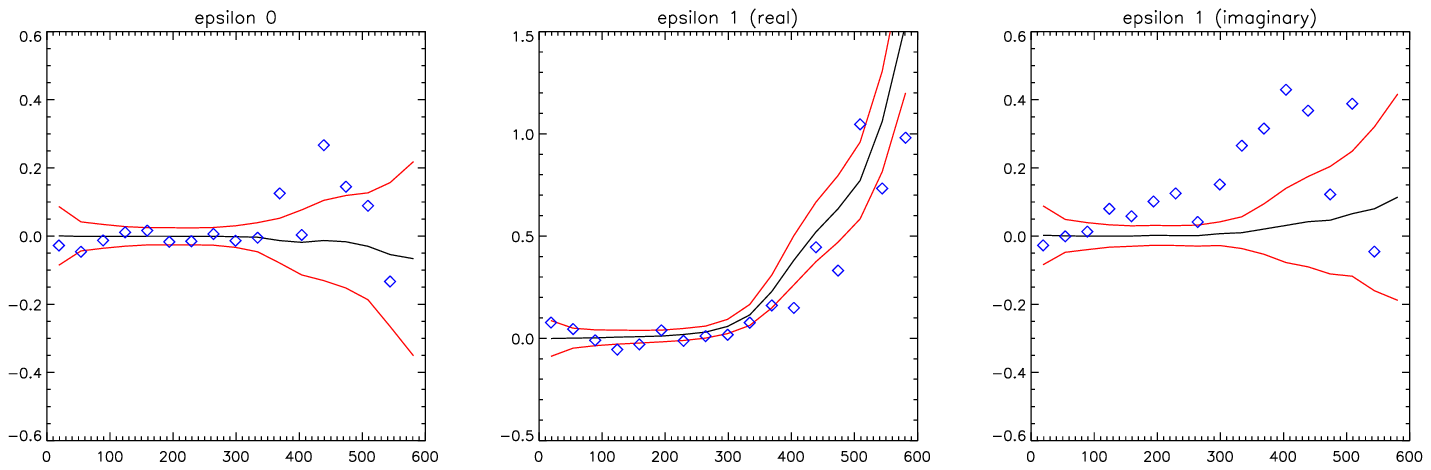,width=18cm}}
  \caption{The analysis of the simulated maps in which (top) the 208 resolved
  sources of the WMAP catalog have been added and (bottom) where the 700 point
  sources (resolved and unresolved) of the V band have been added. In both cases,
  the signal of $\hbox{Re}(\varepsilon_1$ is not affected while the signal
  of $\eo$ and $\hbox{Im}(\varepsilon_1)$ exhibit patterns that are similar
  to the ones obtained in our analysis of the V and W band on the same
  angular scales.}
  \label{fig_final}
\end{figure}

\vskip0.5cm
\noindent{\bf Acknowledgements:} Some of the results of this article
have been derived using the HEALPix package~\cite{package}. We thank
Y. Mellier and R. Stompor for discussions.

\appendix
\section{Integrals over spherical harmonics}\label{appA}

We have evaluated integrals over $n$ spherical harmonics (see
Ref.~\cite{ylm}). When $n=1$ or 2, these integrals are trivial
\begin{eqnarray}
 \int \dd^2\,\bg Y_{\ell
 m}&=&\sqrt{4\pi}\delta_{\ell0}\delta_{m0}\label{A1}\\
 \int \dd^2\,\bg Y_{\la\ma}Y_{\lb\mb}^*&=&\delta_{\la\lb}\delta_{\ma\mb}.\label{A2}
\end{eqnarray}
To go further, one solution is to use the decomposition of the
product of 2 spherical harmonics as
\begin{equation}\label{A3}
 Y_{\la\ma}(\bg)Y_{\lb\mb}(\bg)=\sum_{LM}\sqrt{\frac{(2\la+1)(2\lb+1}{4\pi(2L+1)}}
 C_{\la 0\lb 0}^{L0}C_{\la\ma\lb\mb}^{LM} Y_{LM}(\bg)
\end{equation}
where the $C_{\la\ma\lb\mb}^{LM}$ are the Clebsch-Gordan
coefficients that can be expressed in terms of Wigner $3j$ symbols
as
\begin{equation}\label{A4}
 C_{\la\ma\lb\mb}^{LM}=(-1)^{\la-\lb+M}\sqrt{2L+1}
 \troisj{\la}{\lb}{L}{\ma}{\mb}{-M}
\end{equation}
It is easy to generalize Eq.~(\ref{A3}) to a product of $n$
spherical harmonics
\begin{equation}\label{A5}
 Y_{\la\ma}\ldots Y_{\ell_{n}m_{n}}=\sum_{L_n,M_n}
   \left[
   \sqrt{\frac{4\pi}{2L_n+1}}
   \sum_{L_1\ldots L_{n-1},M_1\ldots M_{n-1}}
   \prod_{i=1}^n\left(\sqrt{\frac{2\ell_i+1}{4\pi}}
   C_{L_{i-1} 0\ell_i 0}^{L_i0}
   C_{L_{i-1}M_{i-1}\ell_i m_i}^{L_iM_i}\right)
   \right]Y_{L_nM_n}.
\end{equation}
We deduce, using Eq.~(\ref{A5}) and the integral (\ref{A2}) that
\begin{eqnarray}
 \int \dd^2\bg Y_{\la\ma}Y_{\lb\mb}Y_{\ell_3m_3}^*&=&
 \sqrt{\frac{(2\la+1)(2\lb+1)}{4\pi(2\ell_3+1)}}
 C_{\la0\lb0}^{\ell_30}
 C_{\la\ma\lb\mb}^{\ell_3m_3}
 \label{A6}\\
 \int \dd^2\bg Y_{\la\ma}Y_{\lb\mb}Y_{\ell_3m_3}Y_{\ell_4 m_4}^*&=&
 \sum_{L,M}
 \sqrt{\frac{(2\la+1)(2\lb+1)(2\ell_3+1)}{(4\pi)^2(2\ell_4+1)}}
 C_{\la0\lb0}^{L0}
 C_{L0\ell_30}^{\ell_40}
 C_{\la\ma\lb\mb}^{LM}
 C_{LM\ell_3m_3}^{\ell_4m_4}.
 \label{A7}
\end{eqnarray}


\end{document}